\documentclass[11pt,a4paper]{article}
\usepackage{amssymb}
\usepackage{amsmath}
\usepackage{graphics}
\usepackage{epsfig}
\usepackage{float}
\usepackage{graphicx}
\usepackage{amsfonts}
\usepackage{amsthm}
\usepackage{newlfont}
\usepackage{color}

\begin{document}

\title{$\Upsilon$ absorption cross sections by nucleons}
\author{Faisal Akram\thanks{%
faisal.chep@pu.edu.pk (corresponding author)}, Bilal Masud\thanks{%
bilalmasud.chep@pu.edu.pk}, Madeeha Nazish\thanks{diha.lifeline@gmail.com} \\
\textit{Center for High Energy Physics, Punjab University, Lahore 54590,
PAKISTAN}}
\maketitle

\begin{abstract}
\noindent The cross sections of $\Upsilon $\ absorption by nucleons are
calculated in meson-baryon exchange model using an effective hadronic
Lagrangian. Calculated cross sections are found to peak near threshold
followed by an increase which tends to settle at large center of mass
energy. Peak values are found in range about 1.8 to 4 mb when cutoff
parameter of monopole form factor is varied from 1.2 to 1.8 GeV. The results
disagree with pQCD calculations but are consistent with the effective value
of the absorption cross section extracted from p-A scattering data in E772
experiment at $\sqrt{s_{NN}}=39$ GeV. We use this data value to find a best
fit of the cutoff parameter which enables us provide definite predictions
for the values of $\Upsilon $ absorption cross sections by nucleons for
upcoming p-A scattering experiments at RHIC and LHC energies for zero
rapidity $b\overline{b}$ states. The results could be useful in studying the
effect of $\Upsilon $ absorption by nucleons at its production stage in
relativistic heavy-ion collision experiments. We also find that thermal
averaged values of total absorption cross sections are small enough to rule
out the effect of absorption of $\Upsilon $ by the comovers in the hadronic
matter.
\end{abstract}


\bigskip

\noindent PACS number(s): 25.75.-q, 14.40.Pq, 14.20.Dh, 13.75.Lb

\section{Introduction}

Both CMS and ALICE collaborations has reported the suppression of the ground and
excited states of $\Upsilon $ in Pb-Pb collisions with respect to their
scaled yield in pp collisions at the same per-nucleon-pair energy \cite%
{cms2013}. The state $\Upsilon (1S)$ is least and $\Upsilon (3S)$ is most
suppressed. This result of increased suppression of less strongly bound
states of $b\overline{b}$ system is consistent with the original idea of
Matsui and Satz \cite{Matsui1986} in which the states of higher radii are
affected before the smaller one due to color Debye screening in quark-gluon
plasma (QGP). However, this observation cannot be regarded as a direct
signal for the existence of the QGP as there are other known mechanism, which
can also affect the production of heavy quarkonia in relativistic heavy-ion
collision experiments. The recombination of $Q\overline{Q}$ pairs due to
heavy quark stopping in deconfined state \cite{regen} and statistical
recombination of heavy quarks can enhance the production rates \cite{regens}%
, whereas interaction with the comovers, mostly $\pi $, $\rho $ mesons and
nucleons, in the hadronic matter can cause dissociation of $J/\psi $ and $%
\Upsilon $. Thus, we require the knowledge of the cross sections of these
processes in order to disentangle the suppression caused by QGP. In Ref.
\cite{Lin2001} the cross section of $\Upsilon $ absorption by $\pi $ and $%
\rho $ mesons are calculated using meson-exchange model. Calculated values
of cross sections $\sigma _{\pi \Upsilon }$ and $\sigma _{\rho \Upsilon }$
are about 8 mb and 1 mb, respectively near threshold energies. However,
their thermal averaged values are very small due large kinetic threshold of
the processes. It is, therefore, claimed in \cite{Lin2001} that the
absorption of $\Upsilon $ by comoving hadrons is unlikely to be important in
heavy-ion collisions. The cross sections of these processes are also
calculated in Ref. \cite{swan2002} using quark potential model.
Corresponding peak values of $\sigma _{\pi \Upsilon } $ and $\sigma _{\rho
\Upsilon }$ are about 0.05 mb and 0.15 mb, respectively. These values are
not only much smaller than those given in \cite{Lin2001}, but also give
entirely different value of the ratio $\sigma _{\pi \Upsilon }/\sigma _{\rho
\Upsilon }$. However, the calculated values of absorption cross sections of $%
\Upsilon ^{^{\prime }}$ for $\pi $ and $\rho $ mesons in Ref.\cite{swan2002}
are about 5 mb and 3 mb respectively. Although the values obtained from
meson exchange and quark potential models disagree, but either way the
thermal averaged values of the cross sections are small enough to
marginalize the effect of these processes. It is also known that production
rate of $\Upsilon $ can also be affected by the dissociation processes $%
\Upsilon +N\rightarrow \Lambda _{b}+\overline{B}/\overline{B}^{\ast }$,
especially in case of appreciable baryonic chemical potential \cite%
{haglin2000}. Thus the values of the cross sections of these processes are
also required before regarding the effect of dissociation by the comovers to
be negligible for $\Upsilon $. Apart from the effect of interaction with the
comovers, the $\Upsilon $ absorption cross sections by nucleons are also
required to study its production in primordial nucleon-nucleon collisions in
heavy-ion collision experiments. It is known that some of the $\Upsilon $
directly produced by these N-N collisions get absorbed by the other nucleons
in pre-equilibrium stage of QGP \cite{Grand2006}. Thus the values of the
absorption cross sections of $\Upsilon $ by nucleons can significantly
affect the expected yield of $\Upsilon $ for RHIC or LHC. Accordingly in the
present work we present the cross sections of these processes, which are
calculated using meson-baryon exchange model based on the same Hadronic
Lagrangian used earlier to calculate the $\Upsilon $ absorption cross
sections by $\pi $ and $\rho $ mesons in \cite{Lin2001} and $B_{c}$
absorption cross sections by nucleons in Ref. \cite{Akram2012}.

Charmonium states analogous to bottomonium are also affected in QGP due to
color Debye screening. In this case the suppression was first studied in
NA50 \cite{NA50} experiment at CERN and then by PHENIX at BNL \cite%
{PHENIX2007}. Like $\Upsilon$ the observed suppression in $J/\psi$ could be
the blend of different effects occurring in the deconfined and the hadronic
states of the matter. Absorption cross sections of $J/\psi $ by light mesons
and nucleons, which are required to study the effect interaction with the
comovers, are rather well studied as compared to $\Upsilon $. Empirical
studies of $J/\psi $ photo-production from nucleons/nuclei \cite{Hufner1998}
and $J/\psi $ production from nucleon-nucleus \cite{Kharzeev1997}
interactions show that $J/\psi $-nucleon cross sections range from $\sim 1$
mb to $\sim 7$ mb. Theoretically, these cross sections have been calculated
using perturbative QCD \cite{Kharzeev1994}, QCD\ sum-rule approach \cite%
{sum-rule}, quark potential models \cite{quark models} and meson-baryon
exchange models based on the hadronic Lagrangian having SU(4) flavor
symmetry \cite{Haglin2000,Lin2000,Liu2001}. The values of the cross sections
obtained from quark potential model and meson-baryon exchange models tend to
agree with the empirical values, but are much larger than those from pQCD or
QCD sum-rule approach. The higher values of the cross sections suggest that
the absorption of $J/\psi $ by the comovers may play significant role. On
the other hand it is also suggested in Refs. \cite{regen,Braun2000} that at
higher collision energy accessible at LHC, large $c\overline{c}$ pairs
production could result into $J/\psi $ enhancement due to heavy quark
stopping effect in QGP. Recent studies of $J/\psi $ in ALICE (A Large Ion
Collider Experiment) at LHC has hinted at this regeneration effect \cite%
{ALICE2011}. Beside the systems of quarkonia, the experimental studies of
the production of heavy mixed flavor mesons and baryons in QGP are also
suggested in Refs. \cite{Schro2000,CUPS2002,Becattini2005}, which predict
enhanced yield of $B_{c}$ mesons, $\Xi _{bc}$, and $\Omega _{ccc}$ baryons
in QGP. Once again the knowledge of the absorption cross sections by
hadronic comovers is required to interpret the observed signal. $B_{c}$
absorption cross sections by pions, $\rho $ and nucleons have been
calculated in Refs. \cite{Akram2012,Akram2011,Lodhi2007} using meson-baryon
exchange model. These cross sections are found to have values of the order
of few mb. In Ref. \cite{Shaheen2019}, it was found that the effect of interactions with comovers is small but not negligible.

Our paper is organized as follows. In Sec. II we produce the interaction
terms required to study $\Upsilon $ absorption by nucleon using the
effective Hadronic models proposed in \cite{Lin2001,Akram2012}. In Sec. III
we produce the amplitudes of the processes. In Sec. IV, we discuss the
numerical values of different couplings used in the calculation. In Sec. V
we present results of cross sections and compare with those obtained from
pQCD and $\Upsilon$ production data in $p-A$ collisions. Finally, some
concluding remarks are made in Sec. VI.

\section{Interaction Lagrangian}

\noindent The following $\Upsilon $ dissociation processes are studied in
this work using meson-baryon exchange model.
\begin{subequations}
\label{1}
\begin{gather}
N\Upsilon \rightarrow B\Lambda _{b},\ \ N\Upsilon \rightarrow B^{\ast
}\Lambda _{b},  \label{1a} \\
N\Upsilon \rightarrow NB^{\ast }\overline{B},\ N\Upsilon \rightarrow NB%
\overline{B}^{\ast },\ N\Upsilon \rightarrow NB\overline{B},\ N\Upsilon
\rightarrow NB^{\ast }\overline{B}^{\ast }.  \label{1b}
\end{gather}%
To calculate the cross sections of the two-body processes of Eq. \ref{1a},
we require the following effective interaction Lagrangian densities.
\end{subequations}
\begin{subequations}
\label{2}
\begin{eqnarray}
\mathcal{L}_{\Upsilon BB} &=&ig_{\Upsilon BB}\Upsilon ^{\mu }(\overline{B}%
\partial _{\mu }B-\partial _{\mu }\overline{B}B),  \label{2a} \\
\mathcal{L}_{\Upsilon B^{\ast }B^{\ast }} &=&ig_{\Upsilon B^{\ast }B^{\ast
}}[(\partial _{\mu }\overline{B}_{\nu }^{\ast }B^{\ast \nu }-\overline{B}%
_{\nu }^{\ast }\partial _{\mu }B^{\ast \nu })\Upsilon ^{\mu }+(\partial
_{\mu }\Upsilon ^{\nu }\overline{B}_{\nu }^{\ast }-\Upsilon ^{\nu }\partial
_{\mu }\overline{B}_{\nu }^{\ast })B^{\ast \mu }  \notag \\
&&+\overline{B}^{\ast \mu }(\Upsilon ^{\nu }\partial _{\mu }B_{\nu }^{\ast
}-\partial _{\mu }\Upsilon ^{\nu }B_{\nu }^{\ast })],  \label{2b} \\
\mathcal{L}_{\Upsilon \Lambda _{b}\Lambda _{b}} &=&g_{\Upsilon \Lambda
_{b}\Lambda _{b}}\overline{\Lambda }_{b}\gamma ^{\mu }\Lambda _{b}\Upsilon
_{\mu },  \label{2c} \\
\mathcal{L}_{BN\Lambda _{b}} &=&ig_{BN\Lambda _{b}}(\overline{N}\gamma
^{5}\Lambda _{b}B+\overline{B}\overline{\Lambda }_{b}\gamma ^{5}N),
\label{2d} \\
\mathcal{L}_{B^{\ast }N\Lambda _{b}} &=&g_{B^{\ast }N\Lambda _{b}}(\overline{%
N}\gamma _{\mu }\Lambda _{b}B^{\ast \mu }+\overline{B}^{\ast \mu }\overline{%
\Lambda }_{b}\gamma _{\mu }N),  \label{2e}
\end{eqnarray}%
\noindent where,
\end{subequations}
\begin{eqnarray}
B &=&\left(
\begin{array}{cc}
B^{+} & B^{0}%
\end{array}%
\right) ^{T},B_{\mu }^{\ast }=\left(
\begin{array}{cc}
B_{\mu }^{\ast +} & B_{\mu }^{\ast 0}%
\end{array}%
\right) ^{T},  \notag \\
N &=&\left(
\begin{array}{c}
p \\
n%
\end{array}%
\right) .
\end{eqnarray}

\noindent Here a field symbol is represented by the
symbol of the particle which it absorbs. Pseudoscalar-pseudoscalar-vector
meson (PPV) and VVV couplings given in Eqs. \ref{2a} and \ref{2b} respectively are
obtained from the hadronic Lagrangian based on SU(5) gauge symmetry in which
the vector mesons are treated as the gauge particles \cite{Lin2001}. In this model
the coupling constants of the allowed interactions
are expressed in terms of single coupling constant $g$. For the
couplings $g_{\Upsilon BB}$ and$\ g_{\Upsilon B^{\ast }B^{\ast }}$
we have%
\begin{equation}
g_{\Upsilon BB}=g_{\Upsilon B^{\ast }B^{\ast }}=\frac{5g}{4\sqrt{10}}.
\label{3}
\end{equation}

\noindent All the mass terms of the vector mesons, which are added directly in the Lagrangian density, break the gauge symmetry explicitly. Thus, we
expected that the symmetry relations given in Eq. \ref{3} may violate.
Baryon-baryon-pseudoscalar meson (BBP) and baryon-baryon-vector meson (BBV)
couplings given in Eqs. \ref{2c} to \ref{2e} can be obtained from the
following SU(5) invariant Lagrangian \cite{Akram2012,Liu2001,Okubo1975}%
\begin{eqnarray}
\mathcal{L}_{PBB} &=&g_{P}(a\phi ^{\ast \alpha \mu \nu }\gamma ^{5}P_{\alpha
}^{\beta }\phi _{\beta \mu \nu }+b\phi ^{\ast \alpha \mu \nu }\gamma
^{5}P_{\alpha }^{\beta }\phi _{\beta \nu \mu }),  \label{4} \\
\mathcal{L}_{VBB} &=&ig_{V}(c\phi ^{\ast \alpha \mu \nu }\gamma. V_{\alpha
}^{\beta }\phi _{\beta \mu \nu }+d\phi ^{\ast \alpha \mu \nu }\gamma.
V_{\alpha }^{\beta }\phi _{\beta \nu \mu }),  \label{5}
\end{eqnarray}

\noindent where all the indices run from 1 to 5. The explicit expressions of the tensors $P_{\alpha
}^{\beta }$ and $V_{\alpha }^{\beta }$ of pseudoscalar and vector mesons respectively are given in Ref. \cite{Lin2001} and the tensor $\phi
^{\alpha \mu \nu }$, which defines the $J^{P}=\frac{1}{2}^{+}$ baryons belonging to 40-plet states in SU(5) quark model, is given in Ref. \cite{Akram2012}. The
Lagrangian densities of Eqs. \ref{4} and \ref{5} define all possible BBP and BBV couplings in
terms of the constants $(g_{P},a,b)$ and $(g_{V},c,d)$ respectively. For the
coupling constants $g_{BN\Lambda _{b}},~$and $g_{B^{\ast }N\Lambda _{b}}$
given in the Eqs. \ref{2d} and \ref{2e}, we have the following results.
\begin{subequations}
\label{6}
\begin{eqnarray}
g_{BN\Lambda _{b}} &=&g_{DN\Lambda _{c}}=g_{KN\Lambda }=\frac{3\sqrt{6}}{8}%
g_{P}(b-a),  \label{6a} \\
g_{B^{\ast }N\Lambda _{b}} &=&g_{D^{\ast }N\Lambda _{c}}=g_{K^{\ast
}N\Lambda }=\frac{3\sqrt{6}}{8}g_{V}(d-c).  \label{6b}
\end{eqnarray}

\noindent SU(5) flavor symmetry in badly broken due to large variation in
the related quark masses. Thus, these symmetry relations are also expected
to be violated.

\noindent In this work we have also included the PVV coupling of $\Upsilon $
meson due to anomalous parity interaction in order to calculate absorption
cross sections by nucleons. The effective Lagrangian density defining the
anomalous interaction of mesons\ is discussed in \cite{oh2001,wsaction}.
Here, we report the relevant interaction term of the Lagrangian density as
follows:
\end{subequations}
\begin{equation}
\mathcal{L}_{\Upsilon B^{\ast }B}=g_{\Upsilon B^{\ast }B}\varepsilon
_{\alpha \beta \mu \nu }\left( \partial ^{\alpha }\Upsilon ^{\beta }\right)
[\partial ^{\mu }B^{\ast \nu }\overline{B}+B\partial ^{\mu }\overline{B}%
^{\ast \nu }],  \label{an int}
\end{equation}

\noindent where $\varepsilon _{\alpha \beta \mu \nu }$ is totally
antisymmetric Levi-Civita tensor of rank four. The coupling constant $%
g_{\Upsilon B^{\ast }B}$, which has the dimension of GeV$^{-1}$, can be
approximated by $g_{\Upsilon BB}/\overline{M}_{B}$ in heavy quark
mass limit \cite{hquark}, where $\overline{M}_{B}$ is the average mass of $B$
and $B^{\ast }$ mesons.

\noindent To calculate the cross sections of the three-body processes given
in Eq. \ref{1b}, we apply the factorization in which double differential
cross sections of the processes are written in terms of the total cross
sections of the correspoding\ two-body subprocesses $\pi \Upsilon
\rightarrow B^{\ast }\overline{B},\ \pi \Upsilon \rightarrow B\overline{B}%
^{\ast },\ \rho \Upsilon \rightarrow B\overline{B},$ and$\ \rho \Upsilon
\rightarrow B^{\ast }\overline{B}^{\ast }$ with off-shell pions or $\rho $
mesons. The cross sections of $\Upsilon $ absorption by on-shell pion and $%
\rho $ mesons are calculated in Ref. \cite{Lin2001}. The expressions of
amplitudes given in this reference can be used to find the values of the
cross section with off-shell $\pi $ and $\rho $ mesons. In addition we also
require $NN\pi $ and $NN\rho $ couplings to apply the factorization formula.
The explicit expression of the these couplings are given in Ref. \cite%
{Liu2001}. Here we reproduce the corresponding Lagrangian densities in
slightly different notations as follows:%
\begin{eqnarray}
\mathcal{L}_{\pi NN} &=&ig_{\pi NN}\overline{N}\gamma ^{5}\overrightarrow{%
\tau }N\cdot \overrightarrow{\pi }, \\
\mathcal{L}_{\rho NN} &=&g_{\rho NN}\overline{N}\left( \gamma ^{\mu }%
\overrightarrow{\tau }+\frac{\kappa _{\rho }}{2m_{N}}\sigma ^{\mu \nu }%
\overrightarrow{\tau }\partial _{\nu }\right) N\cdot \overrightarrow{\rho }%
_{\mu },
\end{eqnarray}

\noindent where, $\pi $ and $\rho $ vectors represent pion and $\rho $
meson isospin triplet respectively. The values of the couplings $g_{\pi NN}$%
, $g_{\rho NN}$ and the tensor coupling $\kappa _{\rho }$ are known
empirically. The numerical values of these couplings and others, which are
required to calculate the cross sections of the $\Upsilon $ absorption
processes, are discussed in section 4.

\section{$\Upsilon $ absorption cross sections}

\noindent Shown in Fig. 1 are the Feynman diagrams of the processes $%
N\Upsilon \rightarrow B\Lambda _{b}$ and$\ N\Upsilon \rightarrow B^{\ast
}\Lambda _{b}$. Total scattering amplitude of the process $N\Upsilon
\rightarrow B\Lambda _{b}$ is given by%
\begin{equation}
M_{1}=\sum\limits_{i=a,b,c}M_{1i}^{\mu }\varepsilon _{\mu }(p_{2}).
\label{amp1}
\end{equation}

\noindent where the amplitudes of the diagrams (1a), (1b), and (1c) are
given by

\begin{figure}[!h]
\begin{center}
\includegraphics[angle=0,width=0.60\textwidth]{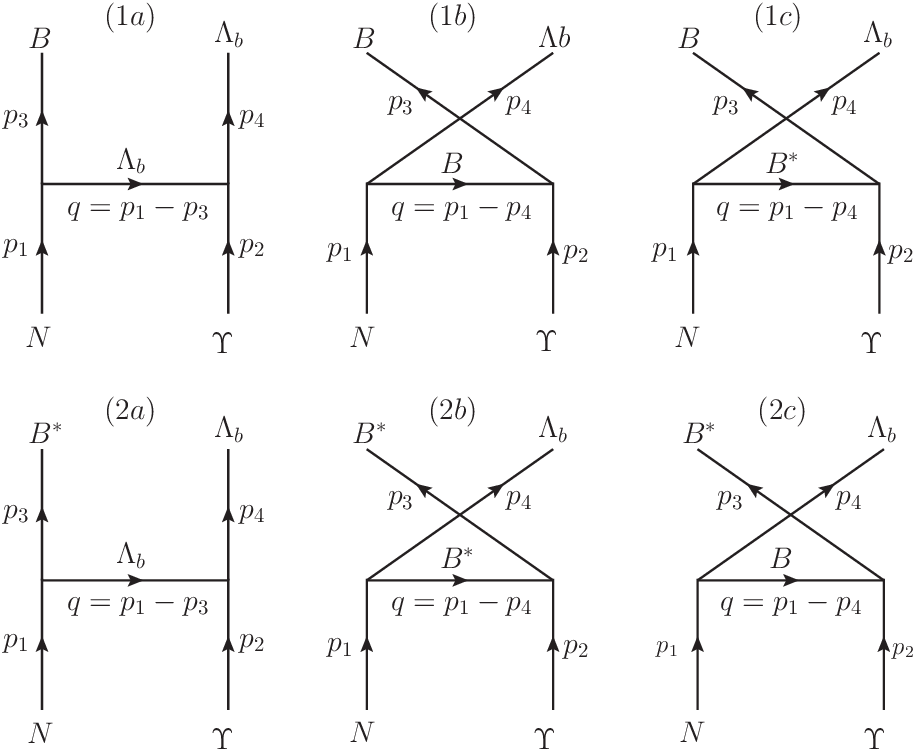}
\end{center}
\caption{Feynman diagrams of $\Upsilon$ absorption processes: $(1)\
N\Upsilon \rightarrow B\Lambda _{b}$ and $(2)\ N\Upsilon \rightarrow B^{\ast
}\Lambda _{b}$}
\label{fig1}
\end{figure}

\begin{subequations}
\label{7}
\begin{eqnarray}
M_{1a}^{\mu } &=&-ig_{\Upsilon \Lambda _{b}\Lambda _{b}}g_{BN\Lambda _{b}}%
\overline{u}_{\Lambda _{b}}(p_{4})\gamma ^{\mu }\left( i\frac{%
(p_{1}-p_{3}).\gamma +m_{\Lambda _{b}}}{t-m_{\Lambda _{b}}^{2}}\right)
\gamma ^{5}u_{N}(p_{1}),  \label{7a} \\
M_{1b}^{\mu } &=&ig_{\Upsilon BB}g_{BN\Lambda _{b}}(2p_{3}-p_{2})^{\mu }%
\frac{i}{u-m_{B}^{2}}\overline{u}_{\Lambda _{b}}(p_{4})\gamma
^{5}u_{N}(p_{1}),  \label{7b} \\
M_{1c}^{\mu } &=&g_{\Upsilon B^{\ast }B}g_{B^{\ast }N\Lambda
_{b}}\varepsilon ^{\alpha \mu \beta \nu }(p_{2})_{\alpha
}(p_{1}-p_{4})_{\beta }\frac{-i}{u-m_{B^{\ast }}^{2}}\left( g_{\nu \lambda }-%
\frac{(p_{1}-p_{4})_{\nu }(p_{1}-p_{4})_{\lambda }}{m_{B^{\ast }}^{2}}\right)
\notag \\
&&\times \overline{u}_{\Lambda _{b}}(p_{4})\gamma ^{\lambda }u_{N}(p_{1}).
\label{7c}
\end{eqnarray}

\noindent And total scattering amplitude of the process $N\Upsilon
\rightarrow B^{\ast }\Lambda _{b}$ is
\end{subequations}
\begin{equation}
M_{2}=\sum\limits_{i=a,b,c}M_{2i}^{\mu \lambda }\varepsilon _{\mu
}(p_{2})\varepsilon _{\lambda }(p_{3}),  \label{amp2}
\end{equation}

\noindent where the partial amplitudes are

\begin{subequations}
\label{8}
\begin{eqnarray}
M_{2a}^{\mu \lambda } &=&-g_{\Upsilon \Lambda _{b}\Lambda _{b}}g_{B^{\ast
}N\Lambda _{b}}\overline{u}_{\Lambda _{b}}(p_{4})\gamma ^{\mu }\left( i\frac{%
(p_{1}-p_{3}).\gamma +m_{\Lambda _{b}}}{t-m_{\Lambda _{b}}^{2}}\right)
\gamma ^{\lambda }u_{N}(p_{1}),  \label{8a} \\
M_{2b}^{\mu \lambda } &=&g_{\Upsilon B^{\ast }B^{\ast }}g_{B^{\ast }N\Lambda
_{b}}[(p_{2}-2p_{3})^{\mu }g^{\nu \lambda }+(p_{2}+p_{3})^{\nu }g^{\mu
\lambda }+(p_{3}-2p_{2})^{\lambda }g^{\mu \nu }]  \notag \\
&&\times \frac{-i}{u-m_{B^{\ast }}^{2}}\left( g_{\sigma \nu }-\frac{%
(p_{1}-p_{4})_{\sigma }(p_{1}-p_{4})_{\nu }}{m_{B^{\ast }}^{2}}\right)
\overline{u}_{\Lambda _{b}}(p_{4})\gamma ^{\sigma }u_{N}(p_{1}),  \label{8b}
\\
M_{2c}^{\mu \lambda } &=&-ig_{\Upsilon B^{\ast }B}g_{BN\Lambda
_{b}}\varepsilon ^{\alpha \mu \beta \lambda }(p_{2})_{\alpha }(p_{3})_{\beta
}\frac{i}{u-m_{B}^{2}}\overline{u}_{\Lambda _{b}}(p_{4})\gamma
^{5}u_{N}(p_{1}).  \label{8c}
\end{eqnarray}

\noindent Using the total amplitudes of these processes, we calculate
unpolarized and isospin averaged cross sections by
\end{subequations}
\begin{equation}
\frac{d\sigma _{i}}{dt}=\frac{I_{i}}{64\pi sp_{i,cm}^{2}}\left\vert
\overline{M}_{i}\right\vert ^{2},\ \ \ \ \ \ \ \ \forall ,~i=1,2  \label{10}
\end{equation}

\begin{figure}[!h]
\begin{center}
\includegraphics[angle=0,width=0.60\textwidth]{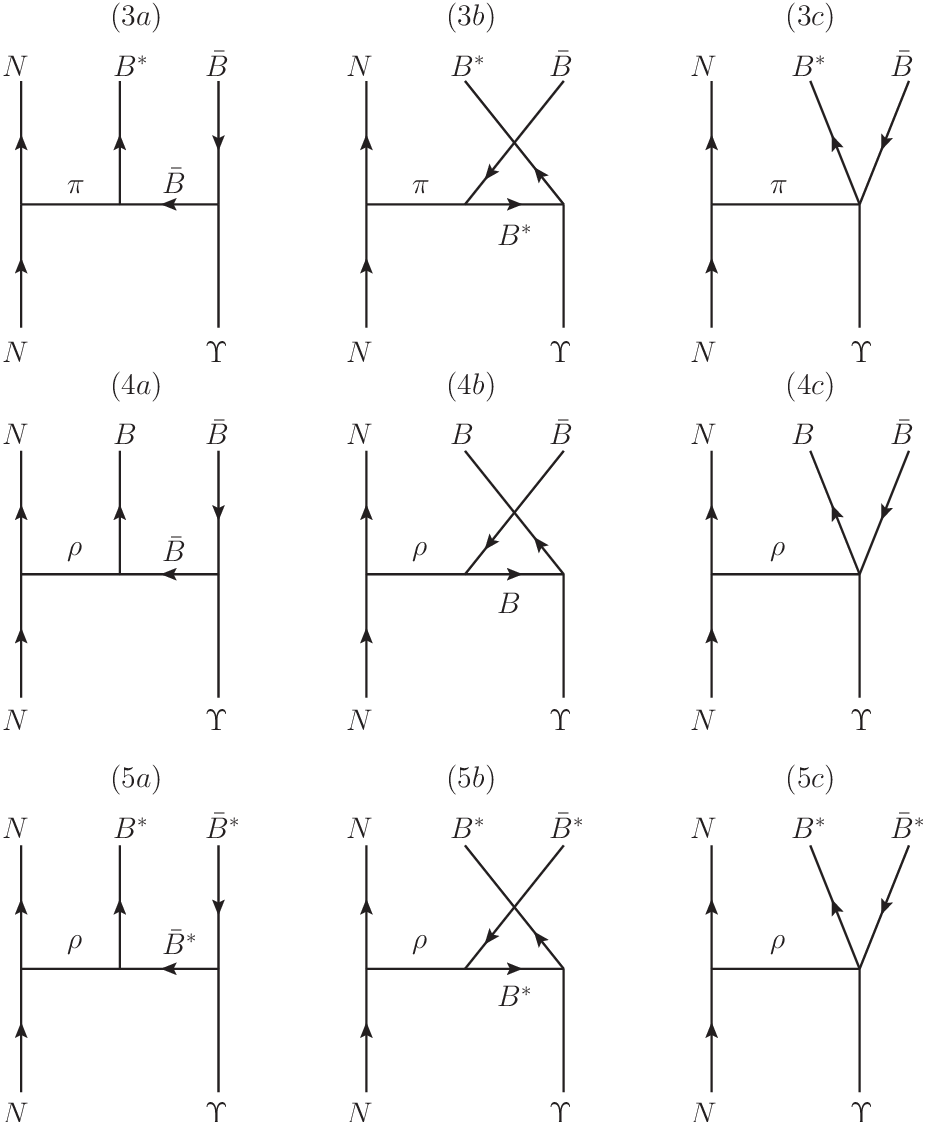}
\end{center}
\caption{Feynman diagrams of $\Upsilon$ absorption processes: $(3)\
N\Upsilon \rightarrow NB^{\ast }\overline{B},\ (4)\ N\Upsilon \rightarrow NB%
\overline{B}$, and $(5)\ N\Upsilon\rightarrow NB^{\ast }\overline{B}^{\ast }$%
}
\label{fig2}
\end{figure}

\noindent where bar over $M_{i}$ stands for averaging (summing) over initial
(final) spins and isospin factors $I_{1}=I_{2}=1$. In Fig. 2 we show the
Feynman diagrams of the three-body processes given in Eq. \ref{1b} excluding
the processes 2 which has same cross section as the processes 1. The
amplitudes of these processes are not given here as they can be obtained
from Eqs. 12 and 13 of Ref. \cite{Liu2001} by simply replacing $D$ and $%
D^{\ast }$ mesons by $B$ and $B^{\ast }$ respectively. The resultant
amplitudes are written in terms of the amplitudes of the subprocesses $\pi
\Upsilon \rightarrow B^{\ast }\overline{B},\ \rho \Upsilon \rightarrow B%
\overline{B},$ and$\ \rho \Upsilon \rightarrow B^{\ast }\overline{B}^{\ast }$%
, which are given separately in Ref. \cite{Lin2001}. The method of Ref. \cite%
{3body} allows us to write the double differential cross sections of the
three-body processes in terms of total cross sections of the corresponding
subprocesses in their c.m frame. Same factorization is also used in Ref.
\cite{Liu2001,Sib2001} for the processes of type $NJ/\Psi \rightarrow
ND^{(\ast )}D^{(\ast )}$. The expressions of double cross sections of the
processes of Eq. \ref{1b} can also be obtained from Eqs. 14 and 15 of Ref.
\cite{Liu2001} by replacing $D$ and $D^{\ast }$ mesons by $B$ and $B^{\ast }$
respectively. Here we reproduce these expressions after correcting a typegraphical error in the Eq. 15 of Ref. \cite{Liu2001} and including the factor 5/4 due to isospin averaging.
\begin{subequations}
\label{d}
\begin{eqnarray}
\frac{d\sigma _{N\Upsilon \rightarrow NB^{\ast }\overline{B}}}{dtds_{1}} &=&%
\frac{(5/4)g_{\pi NN}^{2}}{16\pi ^{2}p_{i}^{2}}k\sqrt{s_{1}}(-t)\frac{F_{\pi
NN}^{2}(t)}{(t-m_{\pi }^{2})^{2}}\sigma _{\pi \Upsilon \rightarrow B^{\ast }%
\overline{B}}(s_{1},t),  \label{da} \\
\frac{d\sigma _{N\Upsilon \rightarrow NB\overline{B}}}{dtds_{1}} &=&\frac{%
(5/4)3g_{\rho NN}^{2}}{32\pi ^{2}p_{i}^{2}}k\sqrt{s_{1}}\frac{F_{\rho
NN}^{2}(t)}{(t-m_{\rho }^{2})^{2}}[4(1+\kappa _{\rho })^{2}(-t-2m_{N}^{2})
\notag \\
&&-\kappa _{\rho }^{2}\frac{(4m_{N}^{2}-t)^{2}}{2m_{N}^{2}}+4(1+\kappa
_{\rho })\kappa _{\rho }(4m_{N}^{2}-t)]\sigma _{\rho \Upsilon \rightarrow B%
\overline{B}}(s_{1},t),  \label{db}
\end{eqnarray}

\noindent where $p_{i}$ is center of mass (c.m) momentum of $N-\Upsilon $ system, $%
t$ is four-momentum square of $\pi $ or $\rho $ meson, and $s_{1}$ and $k$ are the
squared invariant mass and momentum of $\pi -\Upsilon $ in their c.m frame
respectively. The double differential cross section of the process $%
N\Upsilon \rightarrow NB^{\ast }\overline{B}^{\ast }$ can be obtained from
Eq. \ref{db} by replacing $\sigma _{\rho \Upsilon \rightarrow B\overline{B}%
}(s_{1},t)$ with $\sigma _{\rho \Upsilon \rightarrow B^{\ast }\overline{B}%
^{\ast }}(s_{1},t)$. The isospin and spin averaged cross sections $\sigma
_{\pi \Upsilon \rightarrow B^{\ast }\overline{B}}(s_{1},t),$ $\sigma _{\rho
\Upsilon \rightarrow B\overline{B}}(s_{1},t)$, and $\sigma _{\rho \Upsilon
\rightarrow B^{\ast }\overline{B}^{\ast }}(s_{1},t)$ of the subprocesses can
be obtained from Ref. \cite{Lin2001} by replacing the squared mass of pion
and $\rho $ meson by $t$. The form factors $F_{\pi NN}$ and $F_{\rho NN}$ of
the vertices $\pi NN$ and $\rho NN$ are taken in the following monopole form
\end{subequations}
\begin{equation}
F(t)=\frac{\Lambda ^{2}-m^{2}}{\Lambda ^{2}-t},  \label{form1}
\end{equation}

\noindent where $m$ is the mass of exchanged pion or $\rho $ meson and $%
\Lambda $ is the cutoff parameter. Finally the cross sections of the three
body processes can be obtained by integrating Eqs. \ref{da} and \ref{db}.
The limits of integration are given in Ref. \cite{3body}.

\section{Numerical values of the couplings and form factors}

\noindent The values of the couplings $g_{\Upsilon BB}=g_{\Upsilon B^{\ast
}B^{\ast }}=13.3$ are obtained using vector meson dominance (VMD) model in
Ref. \cite{Lin2001}. The BBV coupling $g_{\Upsilon \Lambda _{b}\Lambda _{b}}
$ can be obtained by applying the current conservation conditions on the the
amplitude $M_{1}$:%
\begin{equation}
\sum\limits_{i}M_{1i}^{\mu }(p_{2})_{\mu }=0,
\end{equation}

\noindent in the limit of zero masses. The anomalous term already vanishes in this limit and for the remaining terms we have
\begin{subequations}
\begin{equation}
\sum\limits_{i}M_{1i}^{\mu }(p_{2})_{\mu }=g_{BN\Lambda _{b}}(g_{\Upsilon
BB}-g_{\Upsilon \Lambda _{b}\Lambda _{b}})\overline{u}_{\Lambda
_{b}}(p_{4})\gamma ^{5}u_{N}(p_{1})=0.
\end{equation}

\noindent Thus,
\end{subequations}
\begin{equation}
g_{\Upsilon BB}=g_{\Upsilon B^{\ast }B^{\ast }}=g_{\Upsilon \Lambda
_{b}\Lambda _{b}}=13.3.  \label{vmd}
\end{equation}%
The couplings $g_{BN\Lambda _{b}}$ and $g_{B^{\ast }N\Lambda _{b}}$ can be
fixed by using SU(5) symmetry relations $g_{KN\Lambda }=g_{DN\Lambda
_{c}}=g_{BN\Lambda _{b}}$ and $g_{K^{\ast }N\Lambda }=g_{D^{\ast }N\Lambda
_{c}}=g_{B^{\ast }N\Lambda _{b}}$ given in Eqs. \ref{6} and the empirical
values of the couplings $g_{KN\Lambda }$ and $g_{K^{\ast }N\Lambda }$ given
in Ref. \cite{Stoks1999}. In this way we obtain the following results.%
\begin{equation}
g_{BN\Lambda _{b}}=13.1,\ \ \ \ \ \ \ \ g_{B^{\ast }N\Lambda _{b}}=4.3.
\label{13}
\end{equation}%
However, if estimates $\left\vert g_{DN\Lambda _{c}}\right\vert =7.9$ and $%
\left\vert g_{D^{\ast }N\Lambda _{c}}\right\vert =7.5$ obtained from QCD
sum-rule approach \cite{Duraes2001} are used, we obtain%
\begin{equation}
g_{BN\Lambda _{b}}=7.9,\ \ \ \ \ \ \ \ g_{B^{\ast }N\Lambda _{b}}=7.5.
\label{14}
\end{equation}%
Due to significant difference in the values of the couplings $g_{BN\Lambda
_{b}}$ and $g_{B^{\ast }N\Lambda _{b}}$ obtained from empirical values of $%
(g_{KN\Lambda }$,$~g_{K^{\ast }N\Lambda })$ and QCD sum-rule values of $%
(g_{DN\Lambda _{c}}$,$~g_{D^{\ast }N\Lambda _{c}})$, we use both of them
separately to study their effect on the calculated cross sections. The
couplings $g_{\pi NN}=-13.5$ , $g_{\rho NN}=3.25$, and $\kappa _{\rho }=6.1$
are are extracted from the empirical data using meson exchange model in
Refs. \cite{holzen1989,Jans1996}. Two sets of the values of the coupling
constants used in this paper and methods of obtaining them are summarized in
Table 1.

\begin{table}[tbp] \centering%

\begin{tabular}{c|c|c|c|c}
\hline\hline
& \multicolumn{2}{c}{Set 1}  \vline & \multicolumn{2}{c}{Set 2}  \\ \hline
Coupling Constant & Value & Method of derivation & Value & Method of derivation \\ \hline
$g_{\Upsilon BB}, g_{\Upsilon B^{\ast
}B^{\ast }}$ & 13.3 & VMD model & 13.3 &
VMD model \\
$g_{\Upsilon B^*B}$ & 2.51 & Heavy quark symmetry & 2.51 &
Heavy quark symmetry \\
$g_{DN\Lambda _{c}}$ & 13.1 & $g_{KN\Lambda }$, SU(4) & 7.9 & QCD sum-rule
\\
$g_{D^{\ast }N\Lambda _{c}}$ & 4.3 & $g_{K^{\ast }N\Lambda }$, SU(4) & 7.5 &
QCD sum-rule \\
$g_{BN\Lambda _{b}}$ & 13.1 & $g_{KN\Lambda}$, SU(5) & 7.9 & $%
g_{DN\Lambda _{c}}$, SU(5) \\
$g_{B^{\ast }N\Lambda _{b}}$ & 4.3 & $g_{K^{\ast }N\Lambda}$, SU(5) &
7.5 & $g_{D^{\ast }N\Lambda _{c}}$, SU(5) \\
$g_{\Upsilon \Lambda _{b}\Lambda _{b}}
$ & 13.3 & Current-conservation &
13.3 & Current-conservation \\
\hline\hline
\end{tabular}%
\caption{Two sets of the values of the coupling constants used in this paper.}%
\label{table1}%
\end{table}%

\noindent In order to account for the effect of finite size of interacting
hadrons we need to include the form factors. In this work we use following
monopole form factor at all three point vertices except $\pi NN$ and $\rho
NN $ vertices, where we use Eq. \ref{form1}.

\begin{equation}
f_{3}=\frac{\Lambda ^{2}}{\Lambda ^{2}+\overline{q}^{2}},  \label{16}
\end{equation}

\noindent where $\Lambda $ is the cutoff parameter and $\overline{q}^{2}$ is
squared three momentum transfer in c.m frame of N-$\Upsilon$ system. Generally the value of the
cutoff parameter could be different for different interaction vertices and
can be determined empirically or taken from the theories of the interaction of the
quark constituents of the interacting hadrons. Empirical studies show that
the cutoff parameters vary from 1 to 2 GeV for the vertices connecting
light hadrons ($\pi $, $K$, $\rho $, $N$ etc.) \cite{machleid1987}. However,
due to limited information about the scattering data of bottom hadrons, no
empirical values of the related cutoff parameters are known. In this case we
can approximate the cutoff parameters by relating them with the inverse
(rms) size of hadrons. Using this approximation an estimate is provided in
Ref. \cite{Akram2012}, which shows that the related $\Lambda $ can be varied
from 1.2 to 1.8 GeV to study the uncertainty in the cross section due to the
cutoff parameter. Same form factor\ with the cutoff parameter varied over
the same range is also applied in finding the cross sections of the
subprocesses of $\Upsilon $ absorption by $\pi $ and $\rho $ mesons.
However, for $\pi NN$ and $\rho NN$ vertices we apply the form factor of Eq. %
\ref{form1} in which the cutoff parameter $\Lambda _{\pi NN}=1.3$ GeV \cite%
{holzen1989} and $\Lambda _{\rho NN}=1.4$ GeV \cite{Jans1996}.

\section{Results and Discussion}

\noindent Shown in Fig. 3 are the $\Upsilon $ absorption cross sections by
nucleons for the processes given in Eq. \ref{1a}, as a function of total c.m
energy. These cross sections are obtained using the values of couplings
given in set 1. Solid and dashed curves in the figure represent cross
sections without and with form factors. In each plot the cutoff parameter $%
\Lambda =1.2$ and $1.8$ GeV for lower and upper dashed curves respectively.
Both processes are endothermic, having c.m threshold energies 10.90 and
10.95 GeV respectively. The figure shows that the values of the cross
sections of the processes tend to peak near threshold and then decrease
steadily. The peak values with form factor vary from 1 to 2.5 mb and 0.4 to 1 mb for the
processes $N\Upsilon \rightarrow B\Lambda _{b}$ and $N\Upsilon \rightarrow
B^{\ast }\Lambda _{b}$ respectively.

\begin{figure}[h]
\begin{center}
\includegraphics[angle=0,width=0.45\textwidth]{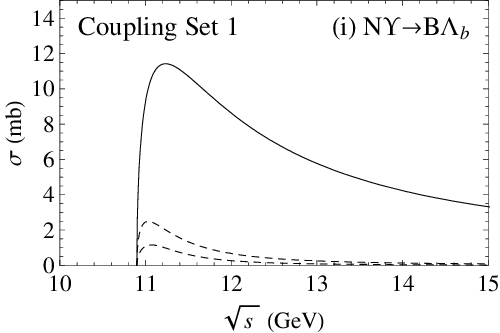} \label{fig3a} %
\includegraphics[angle=0,width=0.45\textwidth]{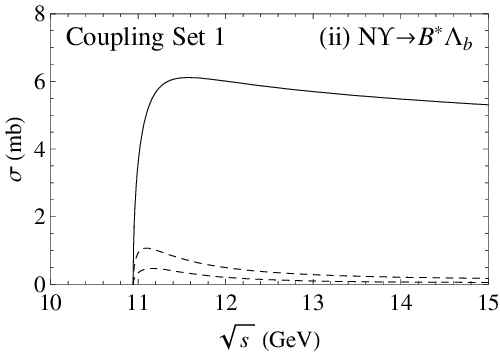} \label{fig3b}
\end{center}
\caption{$\Upsilon $ absorption cross sections of the two body processes
using the values of the couplings given in set 1. Solid and dashed curves
represent cross sections without and with form factor respectively. Lower
and upper dashed curves are with cutoff parameter $\Lambda =1.2$ and $\Lambda
=1.8$ GeV respectively.}
\label{fig3}
\end{figure}

\begin{figure}[h]
\begin{center}
\includegraphics[angle=0,width=0.45\textwidth]{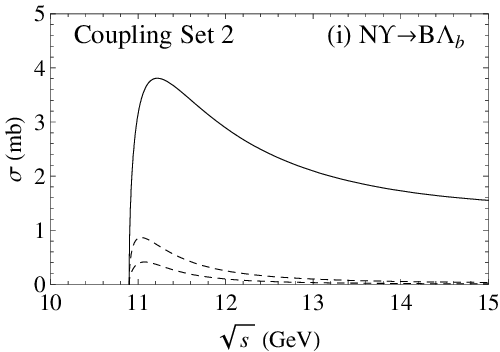} \label{fig4a} %
\includegraphics[angle=0,width=0.45\textwidth]{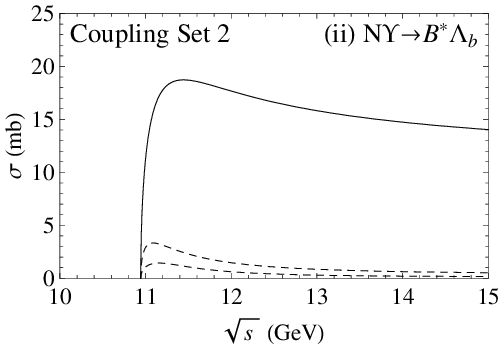} \label{fig4b}
\end{center}
\caption{$\Upsilon $ absorption cross sections of the two body processes
using the values of the couplings given in set 2.}
\label{fig4}
\end{figure}

\noindent In Fig. 4, we present the cross sections of the same processes
using the values of the couplings given in set 2. Again the solid and dashed
curves in the figure represent cross sections without and with form factors
respectively. The plots show similar energy dependence as for set 1 but at
slightly different scale. The peak values near threshold vary from 0.4 to
0.85 mb and 1.4 to 3.4 mb for the first and second process respectively. In
Fig. 5 we show the plots of sum of the cross sections of the two-body
processes as a function of total c.m energy. The peak values of the
accumulated cross sections vary from 1.6 to 3.6 mb and 1.8 to 4.2 mb for
coupling set 1 and 2 respectively. These results show that although the two
coupling sets produce different values of the cross sections for each
process, but the total values almost remains unchanged. In Fig. 6 we show the
plots of the cross sections of the three-body processes $N\Upsilon
\rightarrow NB^{\ast }\overline{B}$ (or $NB\overline{B}^{\ast }$)$,\
N\Upsilon \rightarrow NB\overline{B},\ $and $N\Upsilon \rightarrow NB^{\ast }%
\overline{B}^{\ast }$ for $\Lambda =1.2$ GeV (Left) and $\Lambda =1.8$ GeV
(Right). These processes are also endothermic, having threshold energies $%
10.54$, $10.50$, and $10.59$ GeV respectively. The plots show that the cross
sections of these processes initially increase with energy and then become
almost constant. Furthermore, the cross section of the process $N\Upsilon
\rightarrow NB\overline{B}$ is very small as compared to other processes.
Similar energy dependence is also obtained in case of $J/\psi $ meson \cite%
{Liu2001}. In Fig. 7 we show the plots of the total $\Upsilon $ absorption
cross sections by nucleons as a function of total c.m energy for the two
coupling sets. The plots show that the values of the total cross sections
near threshold are mainly fixed by the two-body processes occurring through
the exchange of bottom mesons and baryons, whereas at large energy the
three-body processes dominate. A close look at the plots of the cross sections
of the processes $N\Upsilon \rightarrow NB^{\ast }\overline{B}$ and$\
N\Upsilon \rightarrow NB^{\ast }\overline{B}^{\ast }$ show that a small
decrease in the former and increase in the latter tend to cancel so that the
resultant total cross sections are relatively stable at large energies as
shown in the Fig. 7.

\begin{figure}[h]
\begin{center}
\includegraphics[angle=0,width=0.45\textwidth]{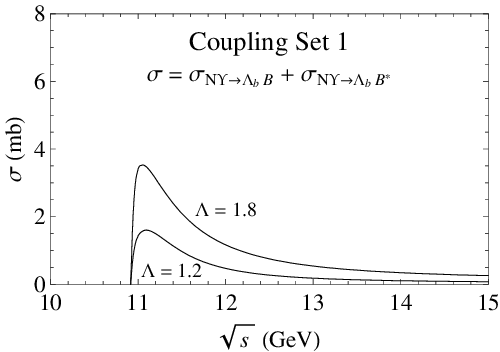} \label{fig5a} %
\includegraphics[angle=0,width=0.45\textwidth]{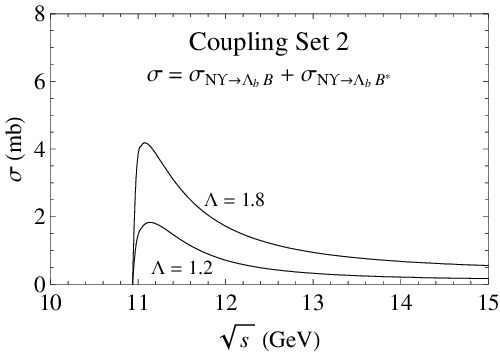} \label{fig5b}
\end{center}
\caption{Sum of cross sections of the two-body processes for coupling set 1
(left) and set 2 (right).}
\label{fig5}
\end{figure}

\begin{figure}[h]
\begin{center}
\includegraphics[angle=0,width=0.45\textwidth]{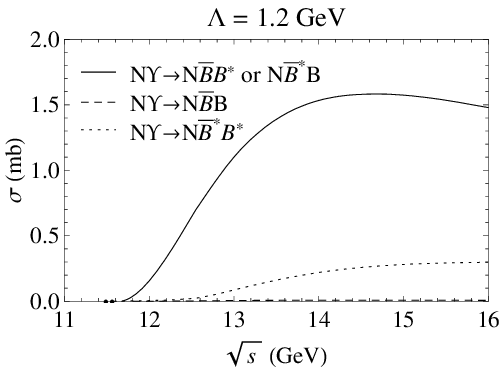} \label{fig6a} %
\includegraphics[angle=0,width=0.45\textwidth]{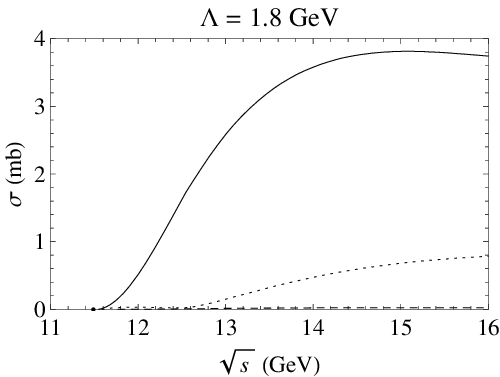} \label{fig6b}
\end{center}
\caption{Cross sections of the three-body processes for $\Lambda=1.2$ GeV (left)
and $\Lambda=1.8$ GeV (right).}
\label{fig6}
\end{figure}

\begin{figure}[h]
\begin{center}
\includegraphics[angle=0,width=0.45\textwidth]{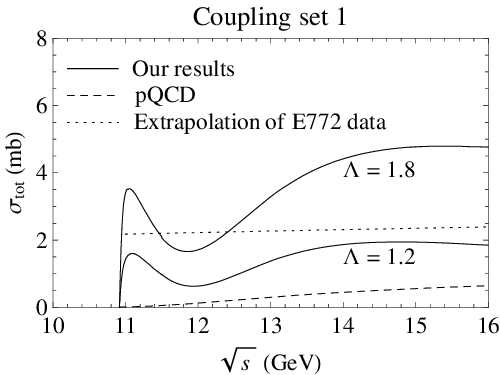} \label{fig7a} %
\includegraphics[angle=0,width=0.45\textwidth]{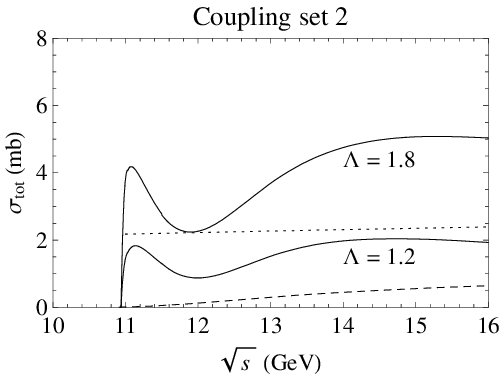} \label{fig7b}
\end{center}
\caption{Total $\Upsilon $ absorption cross section by nucleons for coupling
set 1 (left) and set 2 (right). Dashed curve represents pQCD results of Ref.
\protect\cite{Song2005}.}
\label{fig7}
\end{figure}

Total $\Upsilon $ cross section by nucleon is also calculated in Ref. \cite%
{Song2005} using perturbative QCD \ (pQCD) in next to leading order
approximation. Our results disagree with pQCD in two ways: (i) near
threshold our values of total cross sections are much higher than those of
pQCD and (ii) pQCD results show monotonical increase in the values of cross
sections whereas our results produce a peak near threshold followed by an
increase tending to settle at large c.m energy. Similar disagreement between
pQCD results and other non-perturbative approaches also appears in the case of $%
J/\psi $ cross sections by nucleons as discussed in the section I. Also
mentioned in the section I is that total $\sigma _{N\Upsilon }$ is one of the
input parameter whose value is required to calculate the expected yield of $%
\Upsilon $ in heavy-ion collisions experiments at RHIC or LHC. Experimentally this cross section can be
extracted from the data in p-A collisions or photoproduction of $\Upsilon $.
The available experimental data comes only from E772 experiment \cite{E772}
at Fermi Lab, which show that effective value of the total $\sigma
_{N\Upsilon }$ is about $2.5$ mb at $\sqrt{s_{NN}}=39$ GeV \cite{Beji2003},
which corresponds to $\sqrt{s_{N\Upsilon }}=19.2$ GeV at zero rapidity. This
data value is usually extrapolated to find $\sigma _{N\Upsilon }$ at RHIC
and LHC energies $\sqrt{s_{NN}}=200$ GeV and $5.5\,\ $TeV respectively,
using the model $\sigma _{N\Upsilon }\varpropto \left( \sqrt{s_{NN}}\right)
^{\Delta }$ in which $\Delta$ is taken 0.125: same as for $J/\psi $
absorption by nucleon \cite{Beji2003}. The values of cross sections obtained
from this formula are represented by the dotted curves in Figs. 7 and 8. In
order to provide our estimates at RHIC and LHC energies, we show in Fig. 8
the plots of total $\sigma _{N\Upsilon }$ as a functions of $\sqrt{s_{NN}}$
at the relevant energy scale for different couplings sets and cutoff
parameters. Lower (upper) solid curve represents the values of total $\sigma
_{N\Upsilon }$ using coupling set 1 at $\Lambda =1.2\ $GeV $(\Lambda =1.8$
GeV$)$, whereas dashed curves correspond to coupling set 2. We find that
total $\sigma _{N\Upsilon }$ varies in the ranges 1.0 to 3.0 mb and 1.5 to
3.4 mb at RHIC and LHC energies respectively for either coupling set. In
order to provide definite results, we fit the cutoff parameter $\Lambda $ using
the known value of $\sigma _{N\Upsilon }$ at $\sqrt{s_{NN}}=39$ GeV from
E772 experiment. Fitted values of $\Lambda $ are found to be $1.48$ GeV and $%
1.45$ GeV for coupling set 1 and 2 respectively. These values of cutoff are
then used to produce the plots of the total cross sections represented by
the central curves filled with the green color. We find $\sigma _{N\Upsilon
}=1.8$ mb and $2.2$ mb at $\sqrt{s_{NN}}=200$ GeV and $5.5$ TeV
respectively. These values are lower than 3.1 mb and 4.62 mb respectively,
which are obtained naively from the relation $\sigma _{N\Upsilon }=2.5\left(
\sqrt{s_{NN}}/39\right) ^{0.125}$ and used in Ref. \cite{Grand2006} and
other kinematical studies of time evolution of $\Upsilon $ in RHIC\ and LHC.
Thus we suggest that the results of these kinematical studies should be
reconsidered especially for LHC. Forthcoming results of CMS on $\Upsilon $
production in p-Pb collisions, which will enable us to extract the value of
total $\sigma _{N\Upsilon }$ at $\sqrt{s_{NN}}=5.02$ TeV, could be import in
this regard.

In Fig. 9 we show the plots of thermal averaged values of the total
absorption cross sections as a function of $T$ for both coupling sets. The
colored region between the upper and lower curves represents the uncertainty
due to cutoff parameter. For both coupling sets thermal averaged values are
almost same and much lower than the energy dependent values of cross sections. Large
suppression in the values is due to high kinematic threshold of the
processes. The plots show that at temperature 175 MeV total thermal averaged
value of the absorption cross section is almost 0.4 mb ( 0.2 mb) for $%
\Lambda =1.8$ GeV $(1.2$ GeV$)$. Thus total thermal averaged absorption
cross section of $\Upsilon ,$ after including the value of 0.6 mb for
absorption by $\pi $ and $\rho $ mesons, is not greater than 1 mb at $T=175$
MeV. It is remarked that thermal averaged values of cross sections for $%
T<T_{c}$ are mainly fixed by the values of the cross section at the
energies near-threshold, where only two-body processes dominate. Thus the
role of three-body processes is negligible in defining the thermal averaged
values of the cross sections. In order to make a rough comparison of the
suppression in hadronic matter and the suppression due to QGP, we calculate
upper bound on the decay width or least life time of $\Upsilon $ in hadronic matter by the
following approximate formula \cite{Rapp2010}
\begin{equation}
\Gamma _{\Upsilon }^{HG}=1/\tau _{\Upsilon }^{HG}=\left\langle
n_{tot}^{h}v_{rel}\sigma _{tot}\right\rangle
\end{equation}

\noindent where $n_{tot}$ is total hadronic density which is taken as 0.8 fm$%
^{-3}$ at $T=175$ MeV and $\left\langle v_{rel}\sigma _{tot}\right\rangle =1$
mb. The formula gives $\Gamma _{\Upsilon }^{HG}=16$ MeV, which correspond to
life time $\tau _{\Upsilon }^{HG}=12$ fm. This value is significantly higher
than the life time of $\Upsilon $ due to dissociation in QGP given in Fig. 5
of Ref. \cite{Grand2006}. This suggests that the effect of absorption of $%
\Upsilon $ by comoving hadrons does not play any significant role.

\begin{figure}[h]
\begin{center}
\includegraphics[angle=0,width=0.6\textwidth]{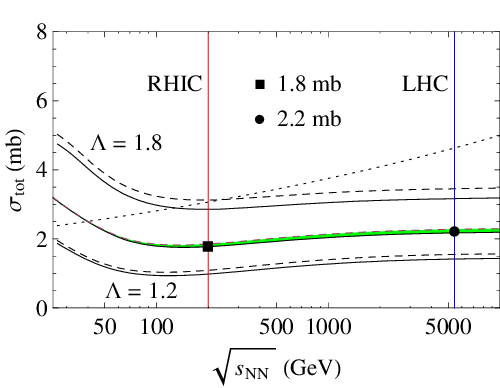} \label{fig8}
\end{center}
\caption{Total $\Upsilon $ absorption cross section by nucleon as a function
$\protect\sqrt{s_{NN}}$ energy for the $\Upsilon $'s produced at zero
rapidity. Conversion between $\protect\sqrt{s_{N\Upsilon }}$ and $\protect%
\sqrt{s_{NN}}$ for zero rapidity particles is made by the formula $%
s_{N\Upsilon }\simeq m_{\Upsilon }\protect\sqrt{s_{NN}}$. Solid and dashed
curves correspond to coupling set 1 and 2 respectively, whereas the dotted
curve represents the values of the cross sections obtained from $\protect%
\sigma _{N\Upsilon }=2.5\left( \protect\sqrt{s_{NN}}/39\right) ^{0.125}$.}
\label{fig8}
\end{figure}

\begin{figure}[h]
\begin{center}
\includegraphics[angle=0,width=0.45\textwidth]{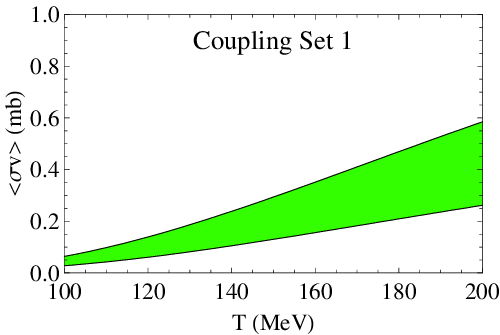} \label{fig9a} %
\includegraphics[angle=0,width=0.45\textwidth]{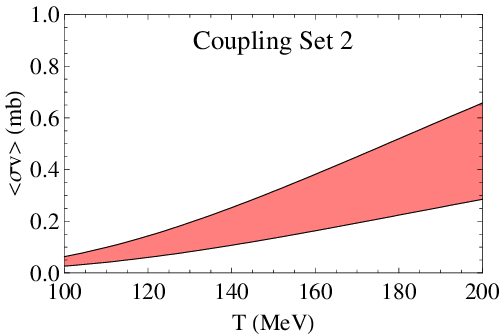} \label{fig9b}
\end{center}
\caption{Thermal averaged values of total absorption cross sections of $%
\Upsilon $ by nucleons as a function of temperature $T$. Colored region
represents the uncertainty due to cutoff parameter.}
\label{fig9}
\end{figure}

\section{Concluding Remarks}

\noindent In this paper, we have calculated $\Upsilon $ absorption cross
sections by nucleons in meson-baryon exchange model using an effective
hadronic Lagrangian. This approach has already been used for calculating
absorption cross sections of $J/\psi $ by light mesons and nucleons, $\Upsilon $ mesons by $\pi$ and $\rho$ mesons, and recently for $B_{c}$ mesons. Main input
quantities required to calculate the cross sections are the effective
couplings and the form factors. In this work the couplings are preferably
fixed empirically, using VMD model, or by vector-current conservation.
Unfortunately at present no information is available for the values of the
couplings $g_{BN\Lambda _{b}}$ and $g_{B^{\ast }N\Lambda _{b}}$ which are required to calculate two-body processes, so we
specify them by the known values of closely related couplings $g_{DN\Lambda
_{c}}$ and $g_{D^{\ast }N\Lambda _{c}}$ as suggested by SU(5) symmetry. On
the bases of two different estimates of the couplings $g_{DN\Lambda _{c}}$
and $g_{D^{\ast }N\Lambda _{c}}$, we calculate the cross sections by
applying two different values of $g_{BN\Lambda _{b}}$ and $g_{B^{\ast }N\Lambda _{b}}$. Our results show that total absorption
cross sections of $\Upsilon $ by nucleons are almost same either way, though
the cross sections of two-body processes are changed some what. So we conclude that
a rigorous study of the couplings $g_{BN\Lambda _{b}}$ and $g_{B^{\ast
}N\Lambda _{b}}$ can further improve our results for these processes. However, our results of total cross sections at high energy are not expected to change as in this regime the cross sections are mainly fixed by three body-processes, which don't depend upon these couplings. Another source of
uncertainty in the cross sections comes from the form factors. In this work
we use the monopole form factors in which the cutoff parameter is to be
fixed empirically or using microscopic theories. However, in the absence of
this information for the bottom mesons, we can constrain cutoff by (rms) size
of interacting hadrons and find that 1.2 to 1.8 GeV is an appropriate range
in which it could be varied to study the effect of its uncertainty on the
cross sections. Same variation in the cutoff is also assumed in other
studies of hadronic cross sections through meson-exchange model. We note that a
known value of the effective cross section extracted from p-A data of E772
experiment can be used to fix the value of cutoff parameter. This allows
us to provide definite predictions of the cross sections at the relevant
energies for the upcoming RHIC and LHC experiments on p-A collisions. Our results could be useful in studying the effect of absorption of $\Upsilon $
by nucleons in pre-equilibrium stage of QGP.

For the energies near
threshold, we find much higher values of total cross sections as compared to
the results obtained from pQCD. However, the corresponding thermal averaged
values of cross section below $T_{c}$ are vary small. Combining these
results with the thermal averaged cross sections of $\Upsilon $ by pion and $%
\rho $ meson, we find that the effect of absorption of $\Upsilon $ by the
comving hadrons after the hadronization is not important.

\section*{Acknowledgement}
FA and MN acknowledge financial support by HEC Pakistan under NRPU grant number 15728.

\end{document}